\documentclass[12pt]{article} 
\usepackage{color}
\usepackage{float}
\usepackage{graphicx}
\usepackage{colortbl}
\usepackage{textcomp}
\usepackage{amssymb}
\usepackage{amsmath}
\usepackage{mathrsfs}
\usepackage{amssymb,amsmath,mathrsfs,latexsym,color,wasysym,graphicx}
\usepackage{enumitem}
\usepackage{multicol}

\newtheorem{theo}{Theorem}[section]
\newtheorem{defn}[theo]{Definition}
\newtheorem{nota}[theo]{Notation}
\newtheorem{exa}[theo]{Example}
\newtheorem{rema}[theo]{Remark}
\newtheorem{coro}[theo]{Corollary}
\newtheorem{lemm}[theo]{Lemma}
\newtheorem{prop}[theo]{Proposition}
\newtheorem{illus}[theo]{Illustration}
\newtheorem{conj}[theo]{Conjecture}

\newcommand{\pf}{\noindent \textit{Proof}:\ }

\def\qed{\ensuremath\square}

\newcommand{\thm}[1]{\begin{theo} #1 \end{theo}}
\renewcommand{\qed}{\hfill{\ensuremath\blacksquare}}

\title{\bf Multi-Strain Age-Structured Dengue Transmission Model: Analysis and Optimal Control \footnote{The first author was funded by the Department of Science and Technology (DOST) of the Philippines and the second author by the Premier Research Institute of Science and Mathematics (PRISM) of MSU-IIT.}}
\author{{\bf Michelle N. Raza$^1$ and Randy L. Caga-anan$^2$}\\
Department of Mathematics and Statistics\\
College of Science and Mathematics\\
MSU-Iligan Institute of Technology\\
Tibanga, Iligan City\\
$^1$michelle.raza@g.msuiit.edu.ph\\
$^2$randy.caga-anan@g.msuiit.edu.ph
}
\date{}

\begin{document}
\maketitle

\begin{abstract}
Dengue is a serious health problem in the Philippines. In 2016, the Department of Health of the country launched a dengue vaccination campaign using Dengvaxia. However, the campaign was mired with controversy and the use of Dengvaxia was banned in the country. This study proposes a mathematical model that represents the  dynamics of the transmission of dengue with its four strains. Considering that the Dengvaxia vaccine was intended to be given only to people aging from 9 to 45 years old, the human population is divided into two age groups: from 9-45 years old and the rest of the population. Using this model and optimal control theory we simulate what could have been the effect of Dengvaxia in the number of dengue cases and then compare this with the other usual dengue intervention strategies. Results show that the best implementation of the usual strategies is better than that of Dengvaxia.  
\end{abstract}


 
{\baselineskip=1.0\baselineskip

\section{Introduction}
Dengue is a  a mosquito-borne viral infection for which there is no specific anti-viral treatment. Dengue virus is transmitted by female mosquitoes mainly of the species \textit{Aedes aegypti} and, to a lesser extent, \textit{Aedes albopictus}. There are four dengue virus serotypes, namely DEN-1, DEN-2, DEN-3, and DEN-4. Infection by one serotype confers life-long immunity to that serotype, but there is no cross-protective immunity to the other serotypes. 

An effective technique used to detect dengue virus and the specific serotype has been developed. This laboratory test involves taking clinical samples and analyzing it through the use of Polymerase Chain Reaction (PCR) \cite{doh2}.  Though the method is very efficient, the laboratory test costs from 4,500 to 7,000 pesos which is expensive for an average Filipino \cite{up}. Data shows that there were a total of 69,088 of dengue cases reported nationwide from January 1 to July 28, 2018 but only 301 cases were confirmed via PCR \cite{doh1}. 

According to the World health Organization, 40 percent of the global population is estimated to be at risk of dengue fever \cite{who2}. Hence, there is a growing public health need for effective preventive interventions against dengue. Dengvaxia is the first dengue vaccine to be licensed. The vaccine was intended to prevent all four dengue types in individuals from 9 to 45 years of age living in endemic areas. In 2017, the manufacturer recommended that the vaccine only be used in people who have previously had a dengue infection, as outcomes may be worsened in those who have not been previously infected.This has caused a scandal in the Philippines where more than 733,000 children were vaccinated regardless of serostatus \cite{who3}. News have exposed deaths due to Dengvaxia and because of this the use of the vaccine is now permanently banned  in the country.

In this paper, we propose a compartmental model describing the  dynamics of dengue transmission considering its four strains. The model is age-structured taking into account the age requirement for the implementation of the Dengvaxia vaccine. Structuring our model this way helps us easily incorporate the control variable for Dengvaxia. There are existing models considering the four strains of dengue but our proposed model have the advantage of focusing only on the number of strains a person have been infected to and do not need any information on what specific strain(s) the person have been infected to. This is mathematically advantageous because the model now needs fewer equations than when one needs to be specific about the strains. This study aims to know how effective Dengvaxia could have been if it was implemented succesfully. This could be done by using optimal control theory on our model. Since Dengvaxia is now banned in the country, we also investigate on the other usual alternatives, like vector control and seeking early medical help and compare these with Dengvaxia.

\section{Model Formulation}

The proposed model considers eight susceptible classes depending on how many times a person was infected with dengue. The researcher considers primary, secondary, tertiary and quaternary infection considering the four serotypes of dengue. But unlike the model in \cite{aguiar}, the specific kind of strain that infects the person  will not matter anymore. The model divides the human population into two compartments: 9-45 years old human population and the rest of the human population taking into account the age implemetantation of the Dengvaxia vaccine. Pandey et al. \cite{pandey} compared SIR model and vector-host model in dengue transmission and finds it that explicitly incorporating the mosquito population may not be necessary in modeling dengue transmission for some populations.

The total population at time $t$, denoted by $N(t)$, is subdivided into eighteen compartments. Age group $a$ consist of individuals of age less than 9 years or greater than 45 years. Age group $b$ consist of individuals of age 9 up to 45 years. Let $i= 1,2,3,4$ and $j=a,b$.  We denote by
\textbf{$S_{ij}(t)$} the number of individuals of age group $j$ who are susceptible to $i$ strain(s) at time $t$;
\textbf{$I_{ij}(t)$} the number of  individuals of age group $j$ who were infected with any $i-1$ strain(s) in the past and is currently (at time $t$) infected with a different strain; and
\textbf{$R_j$} is the number of  individuals of age group $j$ who had been infected by all the strains and has now (at time $t$) completely recovered. It is assumed that all individuals are born susceptible to any of the dengue strains and thus enter the class $S_{4a}$ and a recovered individual from a strain of dengue will have complete immunity to that strain. Humans leave the population through natural death rate and through per-capita death rate due to infection. Humans of age less than 9 years  in the susceptible class $(S_{ia})$ leave the population and move to infected class $(I_{ia})$ or grow older to susceptible class $S_{ib}$. Furthermore, humans of age greater than 45 years in the susceptible class $(S_{ib})$ leave the population and move to infected class $(I_{ib})$ or grow older to susceptible class $S_{ia}$.  In general, any member of the human population remains susceptible with the four strains of dengue and stays in the class $S_{4j}$ for a certain period unless bitten and infected with one strain. Once in the infective class $I_{1j}$, an individual may die or recover from that strain. If recovered from the first infection, susceptibility remains to the other 3 strains and then transfered to $S_{3j}$. If the same individual is infected with another strain, then he will be moved to the class $I_{2j}$. Once recovered from that strain, he is still susceptible to the other two strains and now belongs to the class $S_{2j}$. The cycle repeats until the individual gets infected four times and identified into the class $I_{4j}$. If ever the person survived the fourth infection, he then completely recovers and remains in the class $R_j$.

The parameters used are transmission coefficient ($\alpha_{ij}$), recovery rate ($\beta_{ij}$), birth rate ($\mu$), natural death rate ($\delta$), per-capita death rate through infection ($\gamma_{ij}$), rate of progression from age group $a$ to $b$ ($\epsilon_a$), and rate of progression from age group $b$ to $a$ ($\epsilon_b$). The following figure gives the flow diagram of the dengue transmission model.
\begin{figure}[H]\label{flowchart}
	\begin{center}
		\includegraphics[width=16cm,height=16cm, keepaspectratio]{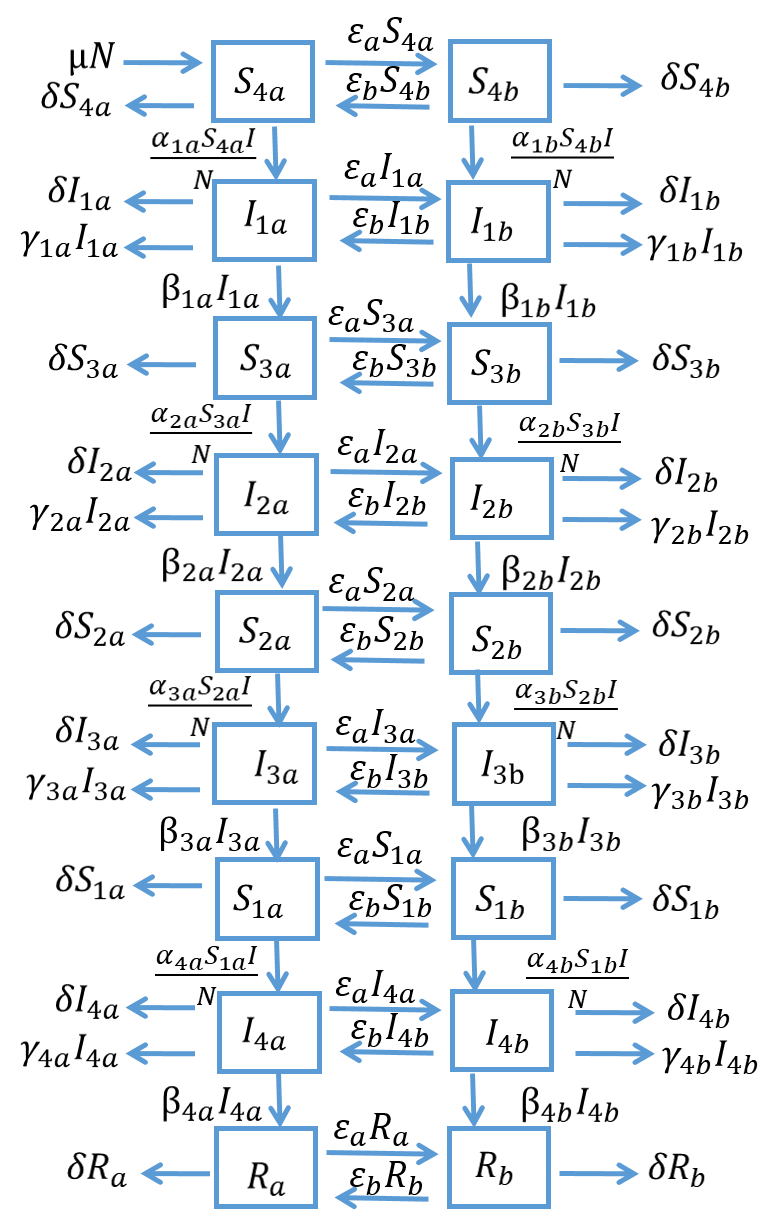}
		\caption{Flow chart of the model}
	\end{center}
\end{figure}

The model can be mathematically described by the following system of 18 differential equations:
\begin{align}\label{main}
\frac{dS_{4a}}{dt}&= \mu N + \epsilon_b S_{4b} - (\epsilon_a + \delta) S_{4a}- \dfrac{\alpha_{1a} S_{4a}I}{N} \notag \\
\frac{dS_{ia}}{dt}&=\epsilon_b S_{ib} + \beta_{(4-i)a}I_{(4-i)a}- (\epsilon_a + \delta) S_{ia}- \dfrac{\alpha_{(5-i)a} S_{ia}I}{N}, \, \, \, \text{for $i=1,2,3$} \notag\\
\frac{dI_{ia}}{dt}&= \dfrac{\alpha_{ia} S_{(5-i)a}I}{N} + \epsilon_b I_{ib} - (\epsilon_a + \delta + \gamma_{ia} + \beta_{ia})I_{ia}, \, \, \, \text{for $i=1,2,3,4$} \notag \\
\frac{dR_a}{dt}&=\beta_{4a}I_{4a} + \epsilon_b R_b - (\delta + \epsilon_a) R_a  \tag{1} \\
\frac{dS_{4b}}{dt}&= \epsilon_a S_{4a} - (\epsilon_b + \delta) S_{4b}- \dfrac{\alpha_{1b} S_{4b}I}{N} \notag \\
\frac{dS_{ib}}{dt}&=\epsilon_a S_{ia} + \beta_{(4-i)b}I_{(4-i)b}- (\epsilon_b + \delta) S_{ib}- \dfrac{\alpha_{(5-i)b} S_{ib}I}{N}, \, \, \, \text{for $i=1,2,3$} \notag \\
\frac{dI_{ib}}{dt}&= \dfrac{\alpha_{ib} S_{(5-i)b}I}{N} + \epsilon_a I_{ia} - (\epsilon_b + \delta + \gamma_{ib} + \beta_{ib})I_{ib}, \, \, \, \text{for $i=1,2,3,4$} \notag \\
\frac{dR_b}{dt}&=\beta_{4b}I_{4b} + \epsilon_a R_a - (\epsilon_b + \delta) R_b \notag 
\end{align}
where $I = \displaystyle\sum_{i=1}^{4} (I_{ia} + I_{ib})$ and $N = \displaystyle\sum_{i=1}^4(S_{ia} + S_{ib} + I_{ia} + I_{ib}) + R_a + R_b $. 
Note that $\dfrac{dN}{dt}= \mu N - \delta N - \displaystyle\sum_{i=1}^4 (\gamma_{ia}I_{ia} + \gamma_{ib}I_{ib}).$
To analyze the model, fractional quantities will be used and this could be done by scaling the population of each class with the total population. The new variables will be as follows:
\begin{equation*}
S_{iA} = \frac{S_{ia}}{N}, I_{iA} = \frac{I_{ia}}{N}, S_{iB} = \frac{S_{ib}}{N}, I_{iB} = \frac{I_{ib}}{N}, R_A = \frac{R_a}{N},  \text{ and } R_B = \frac{R_b}{N},
\end{equation*} for $i=1, 2, 3,4$. 
Note that capital letters are used as subscripts to denote the scaled quantities. Furthermore, observe that 
\begin{equation*}
\sum_{1}^{4}(S_{iA} + S_{iB} + I_{iA} + I_{iB}) + R_A + R_B =1
\end{equation*} or 
\begin{equation*}
S_{4A}= 1- \left(\displaystyle\sum_{i=1}^3(S_{iA})+ \sum_{i=1}^{4}( S_{iB} + I_{iA} + I_{iB}) + R_A + R_B\right).
\end{equation*}
Using this and differentiating each fractional quantities with respect to time, the system (\ref{main}) is reduced to the following system of 17 differential equations:	
{\footnotesize\begin{align}
	\frac{dI_{1A}}{dt}&= \alpha_{1a}I_Nx+ \epsilon_b I_{1B} - (\epsilon_a + \mu + \gamma_{1a} + \beta_{1a})I_{1A} + I_{1A}J \notag \\
	\frac{dS_{iA}}{dt}&=\epsilon_b S_{iB} + \beta_{(4-i)a}I_{(4-i)A}- (\epsilon_a + \mu) S_{iA}- \alpha_{(5-i)a} S_{iA}I_N +  S_{iA}J, \, \, \, \text{for $i=1,2,3$} \notag\\
	\frac{dI_{iA}}{dt}&= \alpha_{ia} S_{(5-i)A}I_N + \epsilon_b I_{iB} - (\epsilon_a + \mu + \gamma_{ia} + \beta_{ia})I_{iA} + I_{iA}J, , \, \, \, \text{for $i=2,3,4$} \notag \\
	\frac{dR_A}{dt}&=\beta_{4a}I_{4A} + \epsilon_b R_B - (\epsilon_a + \mu) R_A + R_AJ \notag\\
	\frac{dS_{4B}}{dt}&=\epsilon_ax - (\epsilon_b + \mu) S_{4B} - \alpha_{1b} S_{4B}I_N + S_{4B}J \tag{2}\\
	\frac{dI_{iB}}{dt}&= \alpha_{1b} S_{(5-i)B}I_N + \epsilon_a I_{iA} - (\epsilon_b + \mu + \gamma_{ib} + \beta_{ib})I_{iB} + I_{iB}J, \, \, \, \text{for $i=1,2,3,4$} \notag\\
	\frac{dS_{iB}}{dt}&=\epsilon_a S_{iA} + \beta_{(4-i)b}I_{(4-i)B}- (\epsilon_b + \mu) S_{iB}- \alpha_{(5-i)b} S_{iB}I_N +  S_{iB}J, \, \, \, \text{for $i=1,2,3$} \notag\\
	\frac{dR_B}{dt}&=\beta_{4B}I_{4B} + \epsilon_a R_A - (\epsilon_b + \mu) R_B + R_B J \notag
	\end{align} }
where $I_N =  \displaystyle\sum_{i=1}^{4} \left(I_{iA} + I_{iB}\right)$ and $J= \sum_{i=1}^{4} (\gamma_{ia} I_{iA} + \gamma_{ib} I_{iB} )$.

This system of equations is epidemiologically and mathematically well-posed on the domain 
\begin{align*}
	D &= \bigg\{ (S_{3A},S_{2A},S_{1A},I_{1A},I_{2A},I_{3A},I_{4A},R_A,  S_{4B},S_{3B},S_{2B},S_{1B},I_{1B},I_{2B},I_{3B}, \\
	&\hspace{1cm}I_{4B},R_B) \in \mathbb{R}_{+}^{17} \, \,\bigg|  \, \, S_{iA} \geq 0, I_{ia} \geq 0, \, \, \,  R_A \geq 0, \, \, \,   S_{iA} \geq 0, \, \, \, I_{iA} \geq 0,\, \, \,  \\
	&\hspace{3cm} R_B \geq 0, \, \, \, \text{where}  \, \, \, i = 1, 2, 3, 4  \, \,  \text{and} \, \, \, \\ & \hspace{2cm}\left(\displaystyle\sum_{i=1}^3(S_{iA} + S_{iB}) + S_{4B} + \sum_{i=1}^{4}( I_{iA} + I_{iB}) + R_a + R_b\right) \leq 1 \bigg\}.
	\end{align*}

The space $\mathbb{R}_{+}^{17}$ denotes the positive orthant in $\mathbb{R}^{17}.$

\subsection{Model Analysis}

In this section, we obtain the disease-free equilibrium point and the reproductive number of the model. A \emph{disease-free equilibrium} (DFE) \label{dfe} is a steady state solution of an epidemic model with all infected variables equal to zero \cite{braun}. Over time, we want to achieve a disease-free state. A threshold that determines whether a disease-free state is achievable or not is the reproductive number $R_0$. The reproductive number  is the expected number of individuals infected by a single infected individual over the duration of the infectious period in a population, which is entirely susceptible \cite{nga}. The reproduction number also gives an idea of which parameters in our model may be significant in our dynamics. The next generation operator approach defined by Diekmann et al. \cite{next} and Driessche and Watmough \cite{nga} will be used to compute $R_0$. 
\thm{Assuming that the initial conditions lie in $D$, the system of equations (4) has a unique solution that exists and remains in $D$ for all time $t \geq 0$.}
\pf Note that the right-hand side of the system of equations (2) is continuous with continuous partial derivatives in $D$, by Cauchy-Lipschitz Theorem, the system of equations (2) has a unique solution. To show that $D$ is forward invariant, note that if $I_{iA} =0$ then $\dfrac{dI_{iA}}{dt} \geq 0$ for $i = 1, 2, 3, 4$. If $S_{iA} =0$ then $\dfrac{dS_{iA}}{dt} \geq 0$ for $i = 1, 2, 3$. If $S_{iB} =0$ then $\dfrac{dS_{iB}}{dt} \geq 0$ for $i = 1, 2, 3, 4$. If $I_{iB} =0$ then $\dfrac{dI_{iB}}{dt} \geq 0$ for $i = 1, 2, 3, 4$. Note also that if $\displaystyle\sum_{i=1}^3(S_{iA} + S_{iB})  + S_{4B} + \sum_{i=1}^{4}( I_{iA} + I_{iB}) + R_A + R_B$ = 1, it follows that 
\begin{align*}
\displaystyle\sum_{i=1}^{3}\frac{dS_{iA}}{dt}+ \displaystyle\sum_{i=1}^{4}\frac{dI_{iA}}{dt}+ \displaystyle\sum_{i=1}^{4}\frac{dS_{iB}}{dt}+ \displaystyle\sum_{i=1}^{4}\frac{dI_{iB}}{dt}  + \frac{dR_A}{dt} + \frac{dR_B}{dt}<0.
\end{align*}
Hence, none of the orbits can leave $D$ and a unique solution exists for all time. \qed

\thm{ The disease-free equilibrium point of the epidemic model (2) is $$\left(0, 0, 0, 0, 0, 0, 0, 0, \dfrac{\epsilon_a}{\epsilon_a +\epsilon_b + \mu}, 0, 0, 0, 0, 0, 0, 0, 0\right).$$}

\pf Set the right-hand side of system (2) to zero and let $I_{iA}=I_{iB}=0$ for $i=1,2,3,4$. Since $\mu >0$ and the DFE point is in $D$, we have $S_{3B} = S_{3A}=S_{2B} = S_{2A} =S_{1B} = S_{1A} = R_A = R_B = 0$.
With these values, we are left with $\epsilon_1 (1 - S_{4B})  - (\epsilon_2 + \mu) S_{4B}=0$. Thus, we have 
$S_{4B}= \dfrac{\epsilon_1}{\epsilon_1 +\epsilon_2 + \mu}$. \qed

\thm{The reproductive number for system (2) is
{\footnotesize \begin{align*} 
R_0&= \alpha_{1a}\left(\dfrac{\epsilon_2 +\mu}{\epsilon_1 +\epsilon_2 + \mu}\right)\left(\dfrac{(\epsilon_2 + \mu + \gamma_{1b} + \beta_{1b}+ \epsilon_1)}{(\epsilon_1 + \mu + \gamma_{1a} + \beta_{1a})(\epsilon_2 + \mu + \gamma_{1b} + \beta_{1b})-\epsilon_1 \epsilon_2 }\right) \\
& \hspace{2cm} +\alpha_{1b}\left(\dfrac{\epsilon_1}{\epsilon_1 +\epsilon_2 + \mu}\right) \left(\dfrac{(\epsilon_1 + \mu + \gamma_{1a} + \beta_{1a} + \epsilon_2) }{(\epsilon_1 + \mu + \gamma_{1a} + \beta_{1a})(\epsilon_2 + \mu + \gamma_{1b} + \beta_{1b})-\epsilon_1 \epsilon_2 }\right),
\end{align*}}
and the system is asymptotically stable if $R_0<1$ and unstable if $R_0>1$.}

\pf
We form the matrices F and V where F is
the matrix of the rates of appearance of new infections and V is the matrix of the rates of transfer
of individuals out of the compartments. Let $X$ be the vector of infected classes and $Y$ be the vector of susceptible and recoverd classes. 

Let $\mathcal{F} (X,Y)$ be the vector of new infection rates (flows from $Y$ to $X$). 
Let $\mathcal{V} (X,Y)$ be the vector of all other rates (not new infection). These rates
include flows from $X$ to $Y$ (for instance, recovery rates), flows within $X$ and flows leaving from the system (for instance, death rates).
The next generation operator formed is $K= FV^{-1}$ where
$F=\left[\dfrac{\partial \mathcal{F}}{\partial X}\right], V=\left[\dfrac{\partial \mathcal{V}}{\partial X}\right]$.
Evaluated at the disease-free equilibrium $(0, 0, 0, 0, 0, 0, 0, 0, S_{4B}^{*}, 0, 0, 0, 0, 0, 0, 0, 0)$ where $S_{4B}^{*}= \dfrac{\epsilon_a}{\epsilon_a +\epsilon_b + \mu}$, this becomes
\begin{align*}
&F(I_{1A}, I_{2A}, \ldots, I_{4B})=\\
&\displaystyle\begin{bmatrix}
\alpha_{1a} x & \alpha_{1a} x & \alpha_{1a} x & \alpha_{1a} x & \alpha_{1a} x & \alpha_{1a} x & \alpha_{1a} x & \alpha_{1a}x\\
0 & 0 & 0 & 0 & 0 & 0 & 0 & 0 \\
0 & 0 & 0 & 0 & 0 & 0 & 0 & 0\\
0 & 0 & 0 & 0 & 0 & 0 & 0 & 0 \\
\alpha_{1b} S_{4B}^{*} & \alpha_{1b} S_{4B}^{*} & \alpha_{1b} S_{4B}^{*} & \alpha_{1b} S_{4B}^{*}& \alpha_{1b} S_{4B}^{*} & \alpha_{1b}S_{4B}^{*} & \alpha_{1b} S_{4B}^{*} & \alpha_{1b} S_{4B}^{*} \\
0 & 0 & 0 & 0 & 0 & 0 & 0 & 0 \\
0 & 0 & 0 & 0 & 0 & 0 & 0 & 0\\
0 & 0 & 0 & 0 & 0 & 0 & 0 & 0 \\
\end{bmatrix}  
\end{align*}
where $x = 1-\dfrac{\epsilon_a}{\epsilon_a +\epsilon_b + \mu} = \dfrac{\epsilon_b + \mu}{\epsilon_a +\epsilon_b + \mu}$.
We also have, 
\begin{align*} 
V(I_{1A}, I_{2A}, \ldots, I_{4B})= \begin{bmatrix}
v_{11}'  & 0 & 0 & 0 & -\epsilon_b & 0 & 0 & 0\\
0 & v_{22}' & 0 & 0 & 0 & -\epsilon_b & 0 & 0\\
0 & 0 & v_{33}' & 0 & 0 & 0 & -\epsilon_b & 0\\
0 & 0 & 0 & v_{44}' & 0 & 0 & 0 & -\epsilon_b\\
-\epsilon_a & 0 & 0 & 0 & v_{55}' & 0 & 0 & 0\\
0 & -\epsilon_a & 0 & 0 & 0 & v_{66}' & 0 & 0\\
0 & 0 & -\epsilon_a & 0 & 0 & 0 & v_{77}' & 0\\
0 & 0 & 0 & -\epsilon_a & 0 & 0 & 0 & v_{88}'\\
\end{bmatrix}
\end{align*} 
where 
\begin{align*}
v_{11}' &= \epsilon_a + \mu + \gamma_{1a} + \beta_{1a} & v_{55}' &= \epsilon_b + \mu + \gamma_{1b} + \beta_{1b} \\ 
v_{22}' &= \epsilon_a + \mu + \gamma_{2a} + \beta_{2a} &v_{66}' &= \epsilon_b + \mu + \gamma_{2b} + \beta_{2b} \\
v_{33}' &= \epsilon_a + \mu + \gamma_{3a} + \beta_{3a} &  v_{77}' &= \epsilon_b + \mu + \gamma_{3b} + \beta_{3b}\\
v_{44}' &= \epsilon_a + \mu + \gamma_{4a} + \beta_{4a} & v_{88}' &= \epsilon_b + \mu + \gamma_{4b} + \beta_{4b}  \\
\end{align*}

Moreover,
\begin{align*} 
V^{-1}(I_{1A}, I_{2A}, \ldots, I_{4B})= \begin{bmatrix}
d_1  & 0 & 0 & 0 & b_1 & 0 & 0 & 0\\
0 & d_2 & 0 & 0 & 0 & b_2 & 0 & 0\\
0 & 0 & d_3 & 0 & 0 & 0 & b_3 & 0\\
0 & 0 & 0 & d_4 & 0 & 0 & 0 & b_4\\
c_1 & 0 & 0 & 0 & d_5 & 0 & 0 & 0\\
0 & c_2 & 0 & 0 & 0 & d_6 & 0 & 0\\
0 & 0 & c_3 & 0 & 0 & 0 & d_7 & 0\\
0 & 0 & 0 & c_4 & 0 & 0 & 0 & d_8\\
\end{bmatrix}
\end{align*}
where
$d_1  = \dfrac{v_{55}'}{v_{11}'v_{55}'-\epsilon_a \epsilon_b}, \,  d_2  = \dfrac{v_{66}'}{ v_{22}'v_{66}'-\epsilon_a \epsilon_b }$, \, $d_3  = \dfrac{v_{77}'}{v_{33}'v_{77}'-\epsilon_a \epsilon_b}, \,  d_4  = \dfrac{v_{88}'}{ v_{44}'v_{88}'-\epsilon_a \epsilon_b }$, \\
$d_5  = \dfrac{v_{11}'}{v_{11}'v_{55}'-\epsilon_a \epsilon_b}, \,  d_6  = \dfrac{-v_{22}'}{v_{22}'v_{66}'-\epsilon_a \epsilon_b}$, \, $d_7  = \dfrac{v_{33}'}{ v_{33}'v_{77}'-\epsilon_a \epsilon_b }, \,  d_8  = \dfrac{v_{44}'}{ v_{44}'v_{88}'-\epsilon_a \epsilon_b }$,\\
$c_1 = \dfrac{\epsilon_a}{v_{11}'v_{55}'-\epsilon_a \epsilon_b}, \,  c_2  = \dfrac{\epsilon_a}{v_{22}'v_{66}'-\epsilon_a \epsilon_b }$, \, $c_3  = \dfrac{\epsilon_a}{v_{33}'v_{77}'-\epsilon_a \epsilon_b}, \,  c_4  = \dfrac{\epsilon_a}{ v_{44}'v_{88}'-\epsilon_a \epsilon_b}$,\\ 
$b_1  = \dfrac{\epsilon_b}{ v_{11}'v_{55}'-\epsilon_a \epsilon_b }, \,  b_2  = \dfrac{\epsilon_b}{ v_{22}'v_{66}'-\epsilon_a \epsilon_b }, \,
b_3  = \dfrac{\epsilon_b}{v_{33}'v_{77}'-\epsilon_a \epsilon_b}, \,  b_4  = \dfrac{\epsilon_b}{v_{44}'v_{88}'-\epsilon_a \epsilon_b }$. 

\vspace{0.2cm}
Now, the next generation matrix $K$ is formed as
\begin{align*}
K=F(I_{1A}, I_{2A}, \ldots, I_{4B})V^{-1}(I_{1A}, I_{2A}, \ldots, I_{4B}).
\end{align*}
Using cofactor expansion to get the eigenvalue of $K$, we have
\begin{align*}
\left|K- \lambda I_8\right| &= (\lambda^6)\left|
\displaystyle\begin{bmatrix}
\alpha_{1a}(1-S_{4B}^{*})(d_1 +c_1) -\lambda & \alpha_{1a}(1-S_{4B}^{*})(d_5 +b_1) \\
\alpha_{1b}S_{4B}^{*}(d_1 +c_1) & \alpha_{1b}S_{4B}^{*}(d_5 +b_1)-\lambda \\
\end{bmatrix}
\right| \\
&= (\lambda^6)\left[-\lambda \alpha_{1a}(1-S_{4B}^{*})(d_1 +c_1) - \lambda \alpha_{1b}S_{4B}^{*}(d_5 +b_1) + \lambda^2  \right] \\
&= (\lambda^7)\left[- \alpha_{1a}x(d_1 +c_1) - \alpha_{1b}S_{4b}^{*}(d_5 +b_1) + \lambda  \right]
\end{align*}
Hence we have, $\lambda= \alpha_{1a}x(d_1 +c_1) +\alpha_{1b}S_{4b}^{*}(d_5 +b_1)$ and $\lambda = 0$ of multiplicity 7. 
Taking the dominant eigenvalue of matrix \emph{$K$}, we get
$R_{0}= \alpha_{1a}(1-S_{4B}^{*})(d_1 +c_1) +\alpha_{1b}S_{4B}^{*}(d_5 +b_1).$ Substituting the values \\ 
$S_{4B}^{*}= \dfrac{\epsilon_a}{\epsilon_a +\epsilon_b + \mu}$, $x= \dfrac{\epsilon_b +\mu}{\epsilon_a +\epsilon_b + \mu}$, $d_1 = \dfrac{v_{55}'}{ v_{11}'v_{55}'-\epsilon_a \epsilon_b}$, $d_5 = \dfrac{v_{11}'}{ v_{11}'v_{55}'-\epsilon_a \epsilon_b }$, $b_1= \dfrac{\epsilon_b}{v_{11}'v_{55}'-\epsilon_a \epsilon_b}$, $c_1= \dfrac{\epsilon_a}{v_{11}'v_{55}'-\epsilon_a \epsilon_b}$, $v_{11}' = \epsilon_a + \mu + \gamma_{1a} + \beta_{1a}$,\\ and  $v_{55}' = \epsilon_b + \mu + \gamma_{1b} + \beta_{1b}$, we finally have
{ \footnotesize \begin{align*}
R_0 &= \alpha_{1a}\left(\dfrac{\epsilon_b +\mu}{\epsilon_a +\epsilon_b + \mu}\right)\left(\dfrac{(\epsilon_b + \mu + \gamma_{1b} + \beta_{1b}+ \epsilon_a)}{(\epsilon_a + \mu + \gamma_{1a} + \beta_{1a})(\epsilon_b + \mu + \gamma_{1b} + \beta_{1b})-\epsilon_a \epsilon_b }\right) \\ & \hspace{1cm} +\alpha_{1b}\left(\dfrac{\epsilon_a}{\epsilon_a +\epsilon_b + \mu}\right) \left(\dfrac{(\epsilon_a + \mu + \gamma_{1a} + \beta_{1a} + \epsilon_b) }{(\epsilon_a + \mu + \gamma_{1a} + \beta_{1a})(\epsilon_b + \mu + \gamma_{1b} + \beta_{1b})-\epsilon_a \epsilon_b }\right).
\end{align*}} \qed

\section{Optimal Control Strategies}
 To mitigate the spread of dengue in the Philippines, three control intervention strategies $u_i(t)$, $i = 1, 2, 3$ were identified to be incorporated in our model to see its best effects in reducing the number of infected individuals at its best implementation with the minimum cost. First is the \textbf{transmission reduction conrol} $u_1(t)$. This consists of holistic and effective methods in fighting mosquito at every stage of their life cycle, such as searching and eliminating breeding grounds, larvicide treatment, using adult mosquito trap, insecticides and thermal fogging during outbreaks. This also encompasses environmental management and sources reduction, self-protection measures which is just strengthening the capability of a person to avoid dengue and health education to provide awareness for their protection against dengue. Next strategy is \textbf{proper medical care control} $u_2(t)$. This means an effort for patients to seek early medication  and immediate reporting to the nearest health care facility as symptoms are observed. With this control, chances of recovery increases specially from severe dengue. Last control is the \textbf{Dengvaxia vaccine} $u_3(t)$. This controversial Dengvaxia vaccine could have been a major breakthrough for dengue prevention in the country, but, the Philippines holds a lot of pending issues against its implementation. Moreover, it has been banned and not allowed for marketing in the Philippines. In this paper, we would like to know what could be the best effect of Dengvaxia (number of infected individuals
 are minimized, while the intervention costs are kept low) if it was properly implemented, that is, if it had been given only to susceptible population from 9-45 years old and to those who have had previous infection. We also investigate on the best possible impact of the other intervention strategies to compare it with Dengvaxia. This study is influenced by a similar study done for tuberculosis in the Philippines in \cite{tb}.

 The dengue state dynamics incorporating the different controls can now be described as follows:
\begin{align}
	\frac{dS_{4a}}{dt}&= \mu N + \epsilon_b S_{4b} - (\epsilon_a + \delta) S_{4a}- (1-u_1(t))\dfrac{\alpha_{1a} S_{4a}I}{N} \notag\\
	\frac{dS_{ia}}{dt}&=\epsilon_b S_{ib} + (1 +u_2(t))\beta_{(4-i)a}I_{(4-i)a}- (\epsilon_a + \delta) S_{ia} \notag\\
	&\hspace{3cm} - (1-u_1(t))\dfrac{\alpha_{(5-i)a} S_{ia}I}{N}, \, \, \, \text{for $i=1,2,3$} \notag \\
	\frac{dI_{ia}}{dt}&= (1-u_1(t))\dfrac{\alpha_{ia} S_{(5-i)a}I}{N} + \epsilon_b I_{ib} - (\epsilon_a + \delta + \gamma_{ia})I_{ia} \notag \\
	&\hspace{3cm}- (1 +u_2(t))\beta_{ia}I_{ia}, \, \, \, \text{for $i=1,2,3,4$} \notag\\
	\frac{dR_a}{dt}&=(1 +u_2(t))\beta_{4a}I_{4a} + \epsilon_b R_b - (\delta + \epsilon_a) R_a \notag \\
	\frac{dS_{4b}}{dt}&= \epsilon_a S_{4a} - (\epsilon_b + \delta) S_{4b}- (1-u_1(t))\dfrac{\alpha_{1b} S_{4b}I}{N} \tag{3} \\
	\frac{dS_{ib}}{dt}&=\epsilon_a S_{ia} + (1 +u_2(t))\beta_{(4-i)b}I_{(4-i)b}- (\epsilon_b + \delta) S_{ib} \notag\\
	&\hspace{3cm}- (1-u_1(t))\dfrac{\alpha_{(5-i)b} S_{ib}I}{N}-u_3S_{ib}, \, \, \, \text{for $i=1,2,3$} \notag \\
	\frac{dI_{ib}}{dt}&= (1-u_1(t))\dfrac{\alpha_{ib} S_{(5-i)b}I}{N} + \epsilon_a I_{ia} - (\epsilon_b + \delta + \gamma_{ib})I_{ib} \notag\\
	&\hspace{3cm} - (1 +u_2(t))\beta_{ib}I_{ib}, \, \, \, \text{for $i=1,2,3$} \notag \\
		\frac{dR_b}{dt}&=(1 +u_2(t))\beta_{4b}I_{4b} + \epsilon_a R_a - (\epsilon_b + \delta) R_b \notag \\
	\dfrac{dN}{dt} &= \mu N - \delta N - \sum_{i=1}^4 (\gamma_{ia}I_{ia} + \gamma_{ib}I_{ib}) - u_3(S_{3b} + S_{2b}+ S_{1b} ) \notag
	\end{align}
Transmission reduction control $u_1(t)$ is applied on all compartments having the transmission rates and proper medical control  $u_2(t)$ is incorporated on all compartments involving recovery rates. The Dengvaxia control  $u_3(t)$ is only applied on age group $b$ since it has restrictions on its implementation to people of age 9-45. Moreover, it is not incorporated to the compartment $S_{4b}$ since it must only be applied to people who have had previous infection.

The controls range is the interval $(0,1)$. The case when $u_i(t) \equiv 0$ means no control effort is employed while the case $u_i(t) \equiv 1$ indicates maximum administration of the control intervention. In this study, the optimal control problem aims to minimize the number of infected individuals with the minimum implementation cost of the control measure(s). It is assumed that the controls are quadratic functions to incorporate nonlinear societal cost associated with the implementation of control measures. This quadratic control is a common form of an objective functional in an optimal control problem. If three controls are considered $u_1(t), u_2(t)$ and $u_3(t)$, the objective functional to be minimized is represented by 
\begin{align*}
J(u_1, u_2, u_3) = \int_{T_0}^{T_1} \left[ I(t) + \dfrac{B_1}{2}u_1^2(t) + \dfrac{B_2}{2}u_2^2(t) + \dfrac{B_3}{2}u_3^2(t)\right]dt 
\end{align*}
where $I(t) = \displaystyle\sum_{i=1}^{4} (I_{ia}(t) + I_{ib}(t))$. The bounds $T_0$ and $T_1$ are taken as 2020 and 2040, respectively, to consider a 20-year dengue intervention program and $B_i$'s, $i=1, 2, 3$, represent the weight constants associated to the relative costs of implementating the respective control strategy. These constants balanced the size and importance of each term in the integrand. In this work, we seek to find optimal controls $u_1^{*}, u_2^{*}$ and $u_3^{*}$ satisfying $$J(u_1^{*}, u_2^{*}, u_3^{*}) = \displaystyle\min_{\Omega} J(u_1, u_2, u_3),$$ where $\Omega = \left\{(u_1, u_2, u_3) \big|\, \,  a \leq u_i(t) \leq b,\, \, \, u_i \in \mathcal{L}^2(2020,2040), \, \, i = 1, 2, 3 \right\}. $ Parameters $a$ and $b$ are the upper and lower bounds of the controls and are assumed to be 0.05 and 0.95, respectively. The weight parameters are set as $B_1 = B_2 = B_3 = 10^6$.

\subsection{Main Theorem}
The goal is to minimize the number of infected individuals and corresponding costs. From Pontryagin's Maximum principle, optimal controls should satisfy the necessary conditions. Pontryagin's Maximum Principle changes into a problem that minimize pointwise a Hamiltonian $H$, with respect to the control. We have $H =  I(t) + \dfrac{B_1}{2}u_1^2(t) + \dfrac{B_2}{2}u_2^2(t) + \dfrac{B_3}{2}u_3^2(t) + \displaystyle\sum_{i=1}^{19}\lambda_i g_i,$ where $g_i$ is the right hand side of the differential equation of the $i$th state variable. Applying Pontryagin's Maximum Principle, we obtain the following theorem.

\thm{There exist optimal controls  $u_1^{*}(t), u_2^{*}(t)$ and $u_3^{*}(t)$ minimizing the objective functional $$\Omega = \left\{(u_1, u_2, u_3) \big|\, \,  a \leq u_i(t) \leq b,\, \, \, u_i \in \mathcal{L}^2(2020,2040), \, \, i = 1, 2, 3 \right\}.$$ Given these optimal solutions, there exist adjoint variables, $\lambda_1(t), \lambda_2(t), \hdots , \lambda_{19}(t)$, which satisfy
	
	{\scriptsize \begin{align*}
		\dfrac{d\lambda_1}{dt} =& \lambda_1\left[(1-u_1(t))\alpha_{1a}\dfrac{I}{N} + \epsilon_1 +\delta \right] - \lambda_2\left[(1-u_1(t))\alpha_{1a}\dfrac{I}{N}\right] - \lambda_{10}\epsilon_1 \\
		\dfrac{d\lambda_2}{dt} =& -1 +\lambda_1\left[(1-u_1(t))\alpha_{1a}\dfrac{S_{4a}}{N}\right]  + \lambda_{19}\gamma_{1a}
		 - \lambda_2\left[(1-u_1(t))\alpha_{1a}\dfrac{S_{4a}}{N} -(\epsilon_1 + \delta) - \gamma_{1a} - (1 +u_2(t))\beta_{1a} \right] \\
		& + \lambda_3 \left[(1-u_1(t))\alpha_{2a}\dfrac{S_{3a}}{N} - (1+u_2(t))\beta_{1a} \right] -\lambda_4\left[ (1-u_1(t))\alpha_{2a}\dfrac{S_{3a}}{N}\right] + \lambda_5\left[ (1-u_1(t))\alpha_{3a}\dfrac{S_{2a}}{N}\right]\\
		& - \lambda_6 \left[ (1-u_1(t))\alpha_{3a}\dfrac{S_{2a}}{N}\right] 
		+ \lambda_7\left[ (1-u_1(t))\alpha_{4a}\dfrac{S_{1a}}{N}\right] - \lambda_8 \left[ (1-u_1(t))\alpha_{4a}\dfrac{S_{1a}}{N}\right] + \lambda_{10}\left[ (1-u_1(t))\alpha_{1b}\dfrac{S_{4b}}{N}\right]\\
		& - \lambda_{11}\left[ (1-u_1(t))\alpha_{1b}\dfrac{S_{4b}}{N} + \epsilon_1 \right] + \lambda_{12}\left[ (1-u_1(t))\alpha_{2b}\dfrac{S_{3b}}{N}\right] - \lambda_{13}\left[ (1-u_1(t))\alpha_{2b}\dfrac{S_{3b}}{N}\right] \\
		& + \lambda_{14}\left[ (1-u_1(t))\alpha_{3b}\dfrac{S_{2b}}{N}\right] - \lambda_{15}\left[ (1-u_1(t))\alpha_{3b}\dfrac{S_{2b}}{N}\right]  + \lambda_{16}\left[ (1-u_1(t))\alpha_{4b}\dfrac{S_{1b}}{N}\right]
		- \lambda_{17}\left[ (1-u_1(t))\alpha_{4b}\dfrac{S_{1b}}{N}\right] \\
		\dfrac{d\lambda_3}{dt} =& \lambda_3\left[(1-u_1(t))\alpha_{2a}\dfrac{I}{N} + \epsilon_1 +\delta \right] - \lambda_4\left[(1-u_1(t))\alpha_{2a}\dfrac{I}{N}\right] - \lambda_{12}\epsilon_1 \\
		\dfrac{d\lambda_4}{dt} =& -1 +\lambda_1\left[(1-u_1(t))\alpha_{1a}\dfrac{S_{4a}}{N}\right]  + \lambda_{19}\gamma_{2a} - \lambda_2 \left[ (1-u_1(t))\alpha_{1a}\dfrac{S_{4a}}{N}\right] + \lambda_3 \left[(1-u_1(t))\alpha_{2a}\dfrac{S_{3a}}{N}\right] \\
		&  -\lambda_4\left[(1-u_1(t))\alpha_{1a}\dfrac{S_{3a}}{N} -(\epsilon_1 + \delta) - \gamma_{2a} - (1 +u_2(t))\beta_{2a} \right] + \lambda_5\left[ (1-u_1(t))\alpha_{3a}\dfrac{S_{2a}}{N}  - (1+u_2(t))\beta_{2a} \right]\\
		& - \lambda_6 \left[ (1-u_1(t))\alpha_{3a}\dfrac{S_{2a}}{N}\right] + \lambda_7\left[ (1-u_1(t))\alpha_{4a}\dfrac{S_{1a}}{N}\right] - \lambda_8 \left[ (1-u_1(t))\alpha_{4a}\dfrac{S_{1a}}{N}\right]  + \lambda_{10}\left[ (1-u_1(t))\alpha_{1b}\dfrac{S_{4b}}{N}\right] \\
		&- \lambda_{11}\left[ (1-u_1(t))\alpha_{1b}\dfrac{S_{4b}}{N} \right] + \lambda_{12}\left[ (1-u_1(t))\alpha_{2b}\dfrac{S_{3b}}{N}\right] - \lambda_{13}\left[ (1-u_1(t))\alpha_{2b}\dfrac{S_{3b}}{N} + \epsilon_1 \right] \\
		& + \lambda_{14}\left[ (1-u_1(t))\alpha_{3b}\dfrac{S_{2b}}{N}\right] - \lambda_{15}\left[ (1-u_1(t))\alpha_{3b}\dfrac{S_{2b}}{N}\right]  + \lambda_{16}\left[ (1-u_1(t))\alpha_{4b}\dfrac{S_{1b}}{N}\right]
		- \lambda_{17}\left[ (1-u_1(t))\alpha_{4b}\dfrac{S_{1b}}{N}\right]\\
		\dfrac{d\lambda_5}{dt} =& \lambda_5\left[(1-u_1(t))\alpha_{3a}\dfrac{I}{N} + \epsilon_1 +\delta \right] - \lambda_6\left[(1-u_1(t))\alpha_{3a}\dfrac{I}{N}\right] - \lambda_{14}\epsilon_1 \\
		\dfrac{d\lambda_6}{dt} =& -1 +\lambda_1\left[(1-u_1(t))\alpha_{1a}\dfrac{S_{4a}}{N}\right]  + \lambda_{19}\gamma_{3a} - \lambda_2 \left[ (1-u_1(t))\alpha_{1a}\dfrac{S_{4a}}{N}\right] + \lambda_3 \left[(1-u_1(t))\alpha_{2a}\dfrac{S_{3a}}{N} \right] \\
		&-\lambda_4\left[ (1-u_1(t))\alpha_{2a}\dfrac{S_{3a}}{N}\right] + \lambda_5\left[ (1-u_1(t))\alpha_{3a}\dfrac{S_{2a}}{N}\right]  - \lambda_6 \left[(1-u_1(t))\alpha_{3a}\dfrac{S_{2a}}{N} -(\epsilon_1 + \delta) - \gamma_{3a} - (1 +u_2(t))\beta_{3a} \right]  \\
		&+ \lambda_7\left[ (1-u_1(t))\alpha_{4a}\dfrac{S_{1a}}{N} - (1+u_2(t))\beta_{3a}\right] - \lambda_8 \left[ (1-u_1(t))\alpha_{4a}\dfrac{S_{1a}}{N}\right]  + \lambda_{10}\left[ (1-u_1(t))\alpha_{1b}\dfrac{S_{4b}}{N}\right]\\
		& - \lambda_{11}\left[ (1-u_1(t))\alpha_{1b}\dfrac{S_{4b}}{N}  \right] + \lambda_{12}\left[ (1-u_1(t))\alpha_{2b}\dfrac{S_{3b}}{N}\right] - \lambda_{13}\left[ (1-u_1(t))\alpha_{2b}\dfrac{S_{3b}}{N}\right] + \lambda_{14}\left[ (1-u_1(t))\alpha_{3b}\dfrac{S_{2b}}{N}\right] \\
		&- \lambda_{15}\left[ (1-u_1(t))\alpha_{3b}\dfrac{S_{2b}}{N} + \epsilon_1 \right] 
		 + \lambda_{16}\left[ (1-u_1(t))\alpha_{4b}\dfrac{S_{1b}}{N}\right]
		- \lambda_{17}\left[ (1-u_1(t))\alpha_{4b}\dfrac{S_{1b}}{N}\right] \\
		\dfrac{d\lambda_7}{dt} =& \lambda_7\left[(1-u_1(t))\alpha_{4a}\dfrac{I}{N} + \epsilon_1 +\delta \right] - \lambda_8\left[(1-u_1(t))\alpha_{4a}\dfrac{I}{N}\right] - \lambda_{16}\epsilon_1 \\
		\dfrac{d\lambda_8}{dt} =& -1 +\lambda_1\left[(1-u_1(t))\alpha_{1a}\dfrac{S_{4a}}{N}\right] - \lambda_9 \left[(1 +u_2(t))\beta_{4a} \right] - \lambda_2 \left[ (1-u_1(t))\alpha_{1a}\dfrac{S_{4a}}{N}\right] + \lambda_3 \left[(1-u_1(t))\alpha_{2a}\dfrac{S_{3a}}{N}\right] \\
		&  -\lambda_4 \left[ (1-u_1(t))\alpha_{2a}\dfrac{S_{3a}}{N}\right] + \lambda_5\left[ (1-u_1(t))\alpha_{3a}\dfrac{S_{2a}}{N} \right]- \lambda_6 \left[ (1-u_1(t))\alpha_{3a}\dfrac{S_{2a}}{N}\right]+ \lambda_7\left[ (1-u_1(t))\alpha_{4a}\dfrac{S_{1a}}{N}\right]\\
		& - \lambda_8 \left[(1-u_1(t))\alpha_{4a}\dfrac{S_{1a}}{N} -(\epsilon_1 + \delta) - \gamma_{4a} - (1 +u_2(t))\beta_{4a} \right] + \lambda_{10}\left[ (1-u_1(t))\alpha_{1b}\dfrac{S_{4b}}{N}\right]\\
		& - \lambda_{11}\left[ (1-u_1(t))\alpha_{1b}\dfrac{S_{4b}}{N} \right] + \lambda_{12}\left[ (1-u_1(t))\alpha_{2b}\dfrac{S_{3b}}{N}\right] - \lambda_{13}\left[ (1-u_1(t))\alpha_{2b}\dfrac{S_{3b}}{N}  \right]  + \lambda_{14}\left[ (1-u_1(t))\alpha_{3b}\dfrac{S_{2b}}{N}\right]\\
		& - \lambda_{15}\left[ (1-u_1(t))\alpha_{3b}\dfrac{S_{2b}}{N}\right] + \lambda_{16}\left[ (1-u_1(t))\alpha_{4b}\dfrac{S_{1b}}{N}\right]
		- \lambda_{17}\left[ (1-u_1(t))\alpha_{4b}\dfrac{S_{1b}}{N} + \epsilon_1\right] + \lambda_{19} \gamma_{4a} \\
		\dfrac{d \lambda_9}{dt} =& \lambda_9(\epsilon_1 + \delta) - \lambda_{18} \epsilon_1 \\
		\dfrac{d\lambda_{10}}{dt} =& \lambda_{10}\left[(1-u_1(t))\alpha_{1b}\dfrac{I}{N} + \epsilon_2 +\delta \right] - \lambda_{11}\left[(1-u_1(t))\alpha_{1b}\dfrac{I}{N}\right] - \lambda_{1}\epsilon_2 \\
					\end{align*} \begin{align*}
		\dfrac{d\lambda_{11}}{dt} =& -1 +\lambda_1\left[(1-u_1(t))\alpha_{1a}\dfrac{S_{4a}}{N}\right]  - \lambda_2 \left[ (1-u_1(t))\alpha_{1a}\dfrac{S_{4a}}{N} + \epsilon_2 \right]  + \lambda_3 \left[(1-u_1(t))\alpha_{2a}\dfrac{S_{3a}}{N}  \right]\\
		& -\lambda_4\left[ (1-u_1(t))\alpha_{2a}\dfrac{S_{3a}}{N}\right] 
		+ \lambda_5\left[ (1-u_1(t))\alpha_{3a}\dfrac{S_{2a}}{N}\right] - \lambda_6 \left[ (1-u_1(t))\alpha_{3a}\dfrac{S_{2a}}{N}\right] \\
		&+ \lambda_7\left[ (1-u_1(t))\alpha_{4a}\dfrac{S_{1a}}{N}\right] - \lambda_8 \left[ (1-u_1(t))\alpha_{4a}\dfrac{S_{1a}}{N}\right] + \lambda_{10}\left[ (1-u_1(t))\alpha_{1b}\dfrac{S_{4b}}{N}\right]\\
		& - \lambda_{11}\left[(1-u_1(t))\alpha_{1b}\dfrac{S_{4b}}{N} -(\epsilon_2 + \delta) - \gamma_{1b} - (1 +u_2(t))\beta_{1b} \right] + \lambda_{12}\left[ (1-u_1(t))\alpha_{2b}\dfrac{S_{3b}}{N} - (1+u_2(t))\beta_{1b}\right]\\
		& - \lambda_{13}\left[ (1-u_1(t))\alpha_{2b}\dfrac{S_{3b}}{N}\right] + \lambda_{14}\left[ (1-u_1(t))\alpha_{3b}\dfrac{S_{2b}}{N}\right] - \lambda_{15}\left[ (1-u_1(t))\alpha_{3b}\dfrac{S_{2b}}{N}\right] \\
		& + \lambda_{16}\left[ (1-u_1(t))\alpha_{4b}\dfrac{S_{1b}}{N}\right]
		- \lambda_{17}\left[ (1-u_1(t))\alpha_{4b}\dfrac{S_{1b}}{N}\right]  + \lambda_{19} \gamma_{1b}\\
		\dfrac{d\lambda_{12}}{dt} =& \lambda_{12}\left[(1-u_1(t))\alpha_{2b}\dfrac{I}{N} + \epsilon_2 +\delta + u_3\right] - \lambda_{13}\left[(1-u_1(t))\alpha_{2b}\dfrac{I}{N}\right] - \lambda_{3}\epsilon_2 + \lambda_{19}u_3 \\
		\dfrac{d\lambda_{13}}{dt} =& -1 +\lambda_1\left[(1-u_1(t))\alpha_{1a}\dfrac{S_{4a}}{N}\right] - \lambda_2 \left[ (1-u_1(t))\alpha_{1a}\dfrac{S_{4a}}{N}\right]  + \lambda_3 \left[(1-u_1(t))\alpha_{2a}\dfrac{S_{3a}}{N}\right] \\
		&-\lambda_4 \left[ (1-u_1(t))\alpha_{2a}\dfrac{S_{3a}}{N} + \epsilon_2 \right] + \lambda_5\left[ (1-u_1(t))\alpha_{3a}\dfrac{S_{2a}}{N}  \right] - \lambda_6 \left[ (1-u_1(t))\alpha_{3a}\dfrac{S_{2a}}{N}\right] \\
		&+ \lambda_7\left[ (1-u_1(t))\alpha_{4a}\dfrac{S_{1a}}{N}\right] - \lambda_8 \left[ (1-u_1(t))\alpha_{4a}\dfrac{S_{1a}}{N}\right] + \lambda_{10}\left[ (1-u_1(t))\alpha_{1b}\dfrac{S_{4b}}{N}\right] - \lambda_{11}\left[ (1-u_1(t))\alpha_{1b}\dfrac{S_{4b}}{N} \right] \\
		&+ \lambda_{12}\left[ (1-u_1(t))\alpha_{2b}\dfrac{S_{3b}}{N} \right] - \lambda_{13}\left[(1-u_1(t))\alpha_{2b}\dfrac{S_{3b}}{N} -(\epsilon_2 + \delta) - \gamma_{2b} - (1 +u_2(t))\beta_{2b} \right]  + \lambda_{19} \gamma_{2b} \\
		& + \lambda_{14}\left[ (1-u_1(t))\alpha_{3b}\dfrac{S_{2b}}{N}  - (1+u_2(t))\beta_{2b}\right] - \lambda_{15}\left[ (1-u_1(t))\alpha_{3b}\dfrac{S_{2b}}{N}\right] \\
		& + \lambda_{16}\left[ (1-u_1(t))\alpha_{4b}\dfrac{S_{1b}}{N}\right]
		- \lambda_{17}\left[ (1-u_1(t))\alpha_{4b}\dfrac{S_{1b}}{N}\right]\\
		\dfrac{d\lambda_{14}}{dt} =& \lambda_{14}\left[(1-u_1(t))\alpha_{3b}\dfrac{I}{N} + \epsilon_2 +\delta + u_3 \right] - \lambda_{15}\left[(1-u_1(t))\alpha_{3b}\dfrac{I}{N}\right] - \lambda_{5}\epsilon_2 + \lambda_{19}u_3\\
		\dfrac{d\lambda_{15}}{dt} =& -1 +\lambda_1\left[(1-u_1(t))\alpha_{1a}\dfrac{S_{4a}}{N}\right] - \lambda_2 \left[ (1-u_1(t))\alpha_{1a}\dfrac{S_{4a}}{N}\right] + \lambda_3 \left[(1-u_1(t))\alpha_{2a}\dfrac{S_{3a}}{N} \right] \\
		&-\lambda_4\left[ (1-u_1(t))\alpha_{2a}\dfrac{S_{3a}}{N}\right] + \lambda_5\left[ (1-u_1(t))\alpha_{3a}\dfrac{S_{2a}}{N}\right] - \lambda_6 \left[ (1-u_1(t))\alpha_{3a}\dfrac{S_{2a}}{N} + \epsilon_2 \right]\\
		& + \lambda_7\left[ (1-u_1(t))\alpha_{4a}\dfrac{S_{1a}}{N}\right] - \lambda_8 \left[ (1-u_1(t))\alpha_{4a}\dfrac{S_{1a}}{N}\right] 
		 + \lambda_{10}\left[ (1-u_1(t))\alpha_{1b}\dfrac{S_{4b}}{N}\right] - \lambda_{11}\left[ (1-u_1(t))\alpha_{1b}\dfrac{S_{4b}}{N}  \right] \\
		&+ \lambda_{12}\left[ (1-u_1(t))\alpha_{2b}\dfrac{S_{3b}}{N}\right] - \lambda_{13}\left[ (1-u_1(t))\alpha_{2b}\dfrac{S_{3b}}{N}\right]- \lambda_{15} \left[(1-u_1(t))\alpha_{3b}\dfrac{S_{2b}}{N} -(\epsilon_2 + \delta) - \gamma_{3b} - (1 +u_2(t))\beta_{3b} \right] \\
		& + \lambda_{14}\left[ (1-u_1(t))\alpha_{3b}\dfrac{S_{2b}}{N}\right] + \lambda_{16}\left[ (1-u_1(t))\alpha_{4b}\dfrac{S_{1b}}{N}  - (1 +u_2(t))\beta_{3b}\right]
		- \lambda_{17}\left[ (1-u_1(t))\alpha_{4b}\dfrac{S_{1b}}{N}\right]  + \lambda_{19} \gamma_{3b} \\
		\dfrac{d\lambda_{16}}{dt} =& \lambda_{16}\left[(1-u_1(t))\alpha_{4b}\dfrac{I}{N} + \epsilon_2 +\delta + u_3\right] - \lambda_{17}\left[(1-u_1(t))\alpha_{4a}\dfrac{I}{N}\right] - \lambda_{7}\epsilon_2 + \lambda_{19}u_3\\
		\dfrac{d\lambda_{17}}{dt} =& -1 +\lambda_1\left[(1-u_1(t))\alpha_{1a}\dfrac{S_{4a}}{N}\right] - \lambda_{18} \left[(1 +u_2(t))\beta_{4b} \right] - \lambda_2 \left[ (1-u_1(t))\alpha_{1a}\dfrac{S_{4a}}{N}\right] + \lambda_3 \left[(1-u_1(t))\alpha_{2a}\dfrac{S_{3a}}{N}\right] \\
		&  -\lambda_4 \left[ (1-u_1(t))\alpha_{2a}\dfrac{S_{3a}}{N}\right] + \lambda_5\left[ (1-u_1(t))\alpha_{3a}\dfrac{S_{2a}}{N} \right]- \lambda_6 \left[ (1-u_1(t))\alpha_{3a}\dfrac{S_{2a}}{N}\right]+ \lambda_7\left[ (1-u_1(t))\alpha_{4a}\dfrac{S_{1a}}{N}\right]\\
		& - \lambda_8 \left[ (1-u_1(t))\alpha_{4a}\dfrac{S_{1a}}{N} + \epsilon_2\right]  + \lambda_{10}\left[ (1-u_1(t))\alpha_{1b}\dfrac{S_{4b}}{N}\right] - \lambda_{11}\left[ (1-u_1(t))\alpha_{1b}\dfrac{S_{4b}}{N} \right] + \lambda_{12}\left[ (1-u_1(t))\alpha_{2b}\dfrac{S_{3b}}{N}\right] \\
		& - \lambda_{13}\left[ (1-u_1(t))\alpha_{2b}\dfrac{S_{3b}}{N}  \right] + \lambda_{14}\left[ (1-u_1(t))\alpha_{3b}\dfrac{S_{2b}}{N}\right] - \lambda_{15}\left[ (1-u_1(t))\alpha_{3b}\dfrac{S_{2b}}{N}\right] +\lambda_{18} \left[(1 +u_2(t))\beta_{4b}\right]\\
		&+ \lambda_{16}\left[ (1-u_1(t))\alpha_{4b}\dfrac{S_{1b}}{N} \right]- \lambda_{17} \left[(1-u_1(t))\alpha_{4b}\dfrac{S_{1b}}{N} -(\epsilon_2 + \delta) - \gamma_{4b} - (1 +u_2(t))\beta_{4b} \right]  + \lambda_{19} \gamma_{4b}\\
			\dfrac{d \lambda_{18}}{dt} =& \lambda_{18}(\epsilon_2 + \delta) - \lambda_{9} \epsilon_2 \\
					\end{align*} \begin{align*}
		\dfrac{d\lambda_{19}}{dt} =& -\lambda_1 \left[\mu + (1-u_1(t))\alpha_{1a}\dfrac{ S_{4a}I}{N^2}\right] +\lambda_2 \left[ (1-u_1(t))\alpha_{1a}\dfrac{ S_{4a}I}{N^2}\right] - \lambda_3 \left[ (1-u_1(t))\alpha_{2a}\dfrac{ S_{3a}I}{N^2}\right]\\
		& + \lambda_4 \left[ (1-u_1(t))\alpha_{2a}\dfrac{ S_{3a}I}{N^2}\right] - \lambda_5 \left[ (1-u_1(t))\alpha_{3a}\dfrac{ S_{2a}I}{N^2}\right]  + \lambda_6 \left[ (1-u_1(t))\alpha_{3a}\dfrac{ S_{2a}I}{N^2}\right]\\
		& - \lambda_7 \left[ (1-u_1(t))\alpha_{4a}\dfrac{ S_{1a}I}{N^2}\right]  + \lambda_8 \left[ (1-u_1(t))\alpha_{4a}\dfrac{ S_{1a}I}{N^2}\right] -\lambda_{10} \left[ (1-u_1(t))\alpha_{1b}\dfrac{ S_{4b}I}{N^2}\right]\\ &+\lambda_{11} \left[ (1-u_1(t))\alpha_{1b}\dfrac{ S_{4b}I}{N^2}\right]
		 - \lambda_{12} \left[ (1-u_1(t))\alpha_{2b}\dfrac{ S_{3b}I}{N^2}\right] + \lambda_{13} \left[ (1-u_1(t))\alpha_{2b}\dfrac{ S_{3b}I}{N^2}\right] +  \lambda_{19}\left[\mu - \delta \right] \\
		   &-\lambda_{14} \left[ (1-u_1(t))\alpha_{3b}\dfrac{ S_{2b}I}{N^2}\right] 
		 + \lambda_{15} \left[ (1-u_1(t))\alpha_{3b}\dfrac{ S_{2b}I}{N^2}\right] - \lambda_{16} \left[ (1-u_1(t))\alpha_{4b}\dfrac{ S_{1b}I}{N^2}\right]  + \lambda_{17} \left[ (1-u_1(t))\alpha_{4b}\dfrac{ S_{1b}I}{N^2}\right] 
		\end{align*} }
	with transversality conditions $\lambda_i(t_f)=0$, for $i = 1, 2,\hdots, 19.$
	Furthermore,
	{\footnotesize \begin{align*}
	u_1^{*}(t) &= \displaystyle \min \left( b, \max \left(a,  \dfrac{-Z}{B_1}\right) \right), \\
	u_2^{*}(t) &= \displaystyle \min \left( b, \max \left(a,   \dfrac{-X}{B_2} \right) \right), \\
	u_3^{*}(t) &=  \displaystyle \min \left( b, \max \left(a, \dfrac{\lambda_{12}S_{3b} + \lambda_{14}S_{2b} + \lambda_{16}S_{1b} + \lambda_{19}\left[S_{3b}+ S_{2b} + S_{1b}\right]}{B_3} \right) \right)
	\end{align*}}
	where
	{\footnotesize \begin{align*}
	Z &= \lambda_1 \dfrac{\alpha_{1a}S_{4a}I}{N} - \lambda_2 \left(\dfrac{\alpha_{1a}S_{4a}I}{N} \right) + \lambda_3 \dfrac{\alpha_{2a}S_{3a}I}{N} - \lambda_4 \left(\dfrac{\alpha_{2a}S_{3a}I}{N} \right)\\
	&\hspace{2cm} \lambda_5 \dfrac{\alpha_{3a}S_{2a}I}{N} - \lambda_6 \left(\alpha_{3a}\dfrac{S_{2a}I}{N} \right) + \lambda_7 \dfrac{\alpha_{4a}S_{1a}I}{N} - \lambda_8 \left(\dfrac{\alpha_{4a}S_{1a}I}{N} \right) \\
	&\hspace{2cm} \lambda_{10} \dfrac{\alpha_{1b}S_{4b}I}{N} - \lambda_{11} \left(\dfrac{\alpha_{1b}S_{4b}I}{N} \right) + \lambda_{12} \dfrac{\alpha_{2b}S_{3b}I}{N} - \lambda_{13} \left(\alpha_{2b}\dfrac{S_{3b}I}{N} \right)\\
	&\hspace{2cm} \lambda_{14} \dfrac{\alpha_{3b}S_{2b}I}{N} - \lambda_{15} \left(\alpha_{3b}\dfrac{S_{2b}I}{N}\right) + \lambda_{16} \dfrac{\alpha_{4b}S_{1b}I}{N} - \lambda_{17} \left(\dfrac{\alpha_{4b}S_{1b}I}{N} \right)
	\end{align*}}
	and
	{\footnotesize\begin{align*}
	X &= (\lambda_3 - \lambda_2)\beta_{1a}I_{1a} + (\lambda_5 - \lambda_4)\beta_{2a}I_{2a}+ (\lambda_7 - \lambda_6)\beta_{3a}I_{3a}  \\
	&\hspace{2cm} (\lambda_{9} - \lambda_{8})\beta_{4a}I_{4a} + (\lambda_{12} - \lambda_{11})\beta_{1b}I_{1b} + (\lambda_{14} - \lambda_{13})\beta_{2b}I_{2b}\\
	&\hspace{4cm} + (\lambda_{16} - \lambda_{15})\beta_{3b}I_{3b}+ (\lambda_{18} - \lambda_{17})\beta_{4b}I_{4b}. 
	\end{align*}}}
\pf The existence of optimal controls $u_1^{*}(t), u_2^*(t),$ and $u_3^*(t)$ such that\\ $J(u_1^{*}(t), u_2^{*}(t), u_3^{*}(t)) = \displaystyle\min_{\Omega} (u_1, u_2, u_3)$, with state system (3) is given by the convexity of the objective functional integrand. By Pontryagin's Maximum Principle, the adjoint equations and transversality conditions can be obtained. Differentiation of the Hamiltonian $H$ with respect to the state variable gives the following system:\\
$\dfrac{d\lambda_1}{dt} = -\dfrac {\partial H}{\partial S_{4a}}, \, \, \, \, \dfrac{d\lambda_2}{dt} = -\dfrac {\partial H}{\partial I_{1a}}, \, \, \, \,\dfrac{d\lambda_3}{dt} = -\dfrac {\partial H}{\partial S_{3a}}, \, \, \dfrac{d\lambda_4}{dt} = -\dfrac {\partial H}{\partial I_{2a}}, \, \, \, \, \\ \dfrac{d\lambda_5}{dt} = -\dfrac {\partial H}{\partial S_{2a}},\, \, \, \, \dfrac{d\lambda_6}{dt} = -\dfrac {\partial H}{\partial I_{3a}}, \, \, \, \, \dfrac{d\lambda_7}{dt} = -\dfrac {\partial H}{\partial S_{1a}}, \, \, \, \,  \dfrac{d\lambda_8}{dt} = -\dfrac {\partial H}{\partial I_{4a}}, \, \,\, \, \\ \dfrac{d\lambda_9}{dt} = -\dfrac {\partial H}{\partial R_{a}}, \, \, \, \, \dfrac{d\lambda_{10}}{dt} = -\dfrac {\partial H}{\partial S_{4b}}, \, \, \, \, \dfrac{d\lambda_{11}}{dt} = -\dfrac {\partial H}{\partial I_{1b}}, \, \, \, \,\dfrac{d\lambda_{12}}{dt} = -\dfrac {\partial H}{\partial S_{3b}}, \, \,\, \,  \\ \dfrac{d\lambda_{13}}{dt} = -\dfrac {\partial H}{\partial I_{2b}}, \, \, \, \,  \dfrac{d\lambda_{14}}{dt} = -\dfrac {\partial H}{\partial S_{2b}},\, \, \, \, \dfrac{d\lambda_{15}}{dt} = -\dfrac {\partial H}{\partial I_{3b}}, \, \, \, \, \dfrac{d\lambda_{16}}{dt} = -\dfrac {\partial H}{\partial S_{1b}}, \, \, \, \, \\ \dfrac{d\lambda_{17}}{dt} = -\dfrac {\partial H}{\partial I_{4b}}, \, \,\, \, \dfrac{d\lambda_{18}}{dt} = -\dfrac {\partial H}{\partial R_{b}}, \, \, \, \, \dfrac{d\lambda_{19}}{dt} = -\dfrac {\partial H}{\partial N},$ \\
with $\lambda_i(t_f)= 0$, for $i=1, 2, \hdots, 19.$

Optimal controls $u_1^{*}(t), u_2^*(t),$ and $u_3^*(t)$ are derived by the following optimality conditions: 

{ \small
	\begin{align*}
	\dfrac {\partial H}{\partial u_1} =& B_1 u_1 +\bigg[\lambda_1 \dfrac{\alpha_{1a}S_{4a}I}{N} - \lambda_2 \left(\dfrac{\alpha_{1a}S_{4a}I}{N}\right) + \lambda_3 \dfrac{\alpha_{2a}S_{3a}I}{N} - \lambda_{17} \left(\dfrac{\alpha_{4b}S_{1b}I}{N} \right) \\
	&- \lambda_4 \left(\dfrac{\alpha_{2a}S_{3a}I}{N}\right) + \lambda_5 \dfrac{\alpha_{3a}S_{2a}I}{N} - \lambda_6 \left(\alpha_{3a}\dfrac{S_{2a}I}{N} \right) + \lambda_7 \dfrac{\alpha_{4a}S_{1a}I}{N} \\
	&- \lambda_8 \left(\dfrac{\alpha_{4a}S_{1a}I}{N} \right) +  \lambda_{10} \dfrac{\alpha_{1b}S_{4b}I}{N} - \lambda_{11} \left(\dfrac{\alpha_{1b}S_{4b}I}{N} \right) + \lambda_{12} \dfrac{\alpha_{2b}S_{3b}I}{N} \\
	& - \lambda_{13} \left(\alpha_{2b}\dfrac{S_{3b}I}{N} \right) + \lambda_{14} \dfrac{\alpha_{3b}S_{2b}I}{N} - \lambda_{15} \left(\alpha_{3b}\dfrac{S_{2b}I}{N} \right) + \lambda_{16} \dfrac{\alpha_{4b}S_{1b}I}{N} = 0 \\
	\dfrac {\partial H}{\partial u_2} &= B_3 u_2  + (\lambda_3 - \lambda_2)\beta_{1a}I_{1a} + (\lambda_5 - \lambda_4)\beta_{2a}I_{2a}+ (\lambda_7 - \lambda_6)\beta_{3a}I_{3a}  \\
	&\hspace{2cm} (\lambda_{9} - \lambda_{8})\beta_{4a}I_{4a} + (\lambda_{12} - \lambda_{11})\beta_{1b}I_{1b} + (\lambda_{14} - \lambda_{13})\beta_{2b}I_{2b}\\
	&\hspace{4cm} + (\lambda_{16} - \lambda_{15})\beta_{3b}I_{3b}+ (\lambda_{18} - \lambda_{17})\beta_{4b}I_{4b} = 0 \\
	\dfrac {\partial H}{\partial u_3} &=B_3 u_3 - \lambda_{12}S_{3b} - \lambda_{14}S_{2b} - \lambda_{16}S_{1b} - \lambda_{19}*(S_{3b} +S_{2b} +S_{1b}) = 0
	\end{align*} }
at $u_1^{*}(t), u_2^*(t),$ and $u_3^*(t)$ on the set $\Omega$. On this set 
\begin{align*}
u_1^{*}(t) &= \dfrac{-Z}{B_1} \\
u_2^{*}(t) &=  \dfrac{-X}{B_2} \\
u_3^{*}(t) &= \frac{1}{B_3}\left[\lambda_{12}S_{3b} + \lambda_{14}S_{2b} + \lambda_{16}S_{1b} + \lambda_{19}(S_{3b} +S_{2b} +S_{1b})\right]
\end{align*} 
where 
{\footnotesize \begin{align*}
	Z &= \lambda_1 \dfrac{\alpha_{1a}S_{4a}I}{N} - \lambda_2 \left(\dfrac{\alpha_{1a}S_{4a}I}{N} + \gamma_{1a} I_{1a}\right) + \lambda_3 \dfrac{\alpha_{2a}S_{3a}I}{N} - \lambda_4 \left(\dfrac{\alpha_{2a}S_{3a}I}{N} + \gamma_{2a} I_{2a}\right)\\
	&\hspace{1cm} \lambda_5 \dfrac{\alpha_{3a}S_{2a}I}{N} - \lambda_6 \left(\alpha_{3a}\dfrac{S_{2a}I}{N} + \gamma_{3a} I_{3a}\right) + \lambda_7 \dfrac{\alpha_{4a}S_{1a}I}{N} - \lambda_8 \left(\dfrac{\alpha_{4a}S_{1a}I}{N} + \gamma_{4a} I_{4a}\right) \\
	&\hspace{1cm} \lambda_{10} \dfrac{\alpha_{1b}S_{4b}I}{N} - \lambda_{11} \left(\dfrac{\alpha_{1b}S_{4b}I}{N} + \gamma_{1b} I_{1b}\right) + \lambda_{12} \dfrac{\alpha_{2b}S_{3b}I}{N} - \lambda_{13} \left(\alpha_{2b}\dfrac{S_{3b}I}{N} + \gamma_{2b} I_{2b}\right)\\
	&\hspace{1cm} \lambda_{14} \dfrac{\alpha_{3b}S_{2b}I}{N} - \lambda_{15} \left(\alpha_{3b}\dfrac{S_{2b}I}{N} + \gamma_{3b} I_{3b}\right) + \lambda_{16} \dfrac{\alpha_{4b}S_{1b}I}{N} - \lambda_{17} \left(\dfrac{\alpha_{4b}S_{1b}I}{N} + \gamma_{4b} \right) \\
	&\hspace{4cm} + \lambda_{19} \left(\sum_{i=1}^4 (\gamma_{ia}I_{ia} + \gamma_{ib}I_{ib})\right),
	\end{align*}}
and
\begin{align*}
X &= (\lambda_3 - \lambda_2)\beta_{1a}I_{1a} + (\lambda_5 - \lambda_4)\beta_{2a}I_{2a}+ (\lambda_7 - \lambda_6)\beta_{3a}I_{3a}  \\
&\hspace{2cm} (\lambda_{9} - \lambda_{8})\beta_{4a}I_{4a} + (\lambda_{12} - \lambda_{11})\beta_{1b}I_{1b} + (\lambda_{14} - \lambda_{13})\beta_{2b}I_{2b}\\
&\hspace{4cm} + (\lambda_{16} - \lambda_{15})\beta_{3b}I_{3b}+ (\lambda_{18} - \lambda_{17})\beta_{4b}I_{4b}. 
\end{align*}
Taking into account the bounds on controls, we obtain the characterization of $u_1^{*}(t), u_2^*(t),$ and $u_3^*(t)$. We have
\begin{align*}
u_1^{*}(t) &= \displaystyle \min \left( b, \max \left(a,\dfrac{-Z}{B_1}\right) \right), \\
u_2^{*}(t) &= \displaystyle \min \left( b, \max \left(a, \dfrac{-X}{B_2}  \right) \right), \\
u_3^{*}(t) &=  \displaystyle \min \left( b, \max \left(a, \frac{1}{B_3}\left[\lambda_{12}S_{3b} - \lambda_{14}S_{2b} - \lambda_{16}S_{1b} - \lambda_{19}*(S_{3b} +S_{2b} +S_{1b}) \right]\right) \right).
\end{align*}
\qed

\subsection{Estimating Parameters}
Pandey et al. \cite{pandey} compared the SIR models and the vector-host models for dengue transmission and found that explicitly incorporating the mosquito population may not be necessary in modeling dengue transmission for some populations. In their paper, by comparing the equilibria of the vector host model and the SIR model they obtained the transmission coefficient $\beta$ in terms of the parameters of the vector host model: $\beta \approx \dfrac{mc^2\beta_H\beta_V}{\mu_V}$ where $m$ is the number of mosquitos per person, $c$ is the biting rate, $\beta_H$ is the mosquito-to-human transmission probability, $\beta_V$ is the human-to-mosquito transmission probability, $\mu_V$ is the mosquito mortality rate and $\beta$ is the composite human to human transmission rate. Note that in our paper, we use $\alpha$ as our transmission coefficient. From \cite{dengue}, we can get the following values: biting rate $c = 1$, mosquito-to-human transmission probability $\beta_H = 0.375$, human-to-mosquito transmission probability $\beta_V = 0.75$, mosquito mortality rate $\mu_V = 0.1$. From \cite{malaria}, we also get the value of the number of mosquito per person $m$ =1.   Substituting these parameter values to $\alpha$, we have $$ \dfrac{mc^2\beta_H\beta_V}{\mu_V} = \dfrac{1(1)^2(0.375)(0.75)}{0.1} = 2.8125  .$$ This now becomes the value for $\alpha_{1a}$ and $\alpha_{1b}$. 

Data from \cite{doh1} reveals that 54 \% of the dengue cases is caused by DENV 3, 25 \% is caused by DENV 1, 18 \% is caused by DENV 2 and 3 \%  of the dengue cases is caused by DENV 4. Assuming that this distribution directly affect the  transmission coefficients for the second, third and fourth infections, we can arrive at the following estimates for the other transmission coefficients: $\alpha_{2a} = 0.6125999\times\alpha_{1a}$, $\alpha_{3a} = 0.195696\times\alpha_{1a}$, $\alpha_{4a} = 0.017496\times\alpha_{1a}$, $\alpha_{2b} = 0.6125999\times\alpha_{1b}$, $\alpha_{3b} = 0.195696\times\alpha_{1b}$, and $\alpha_{4b} = 0.017496\times\alpha_{1b}$. Thus, we have $\alpha_{2a} = 1.7229$, $\alpha_{3a} = 0.550395$, $\alpha_{4a} = 0.0478125$, $\alpha_{2b} = 1.7229$, $\alpha_{3b} = 0.550395$, and $\alpha_{4b} = 0.0478125$.

From Syafruddin and Noorani model of dengue fever in \cite{noorani}, the recovery rate of infected humans is 0.32833. But data from \cite{doh1} shows that the recovery of people of age less than nine years old, where most dengue cases occur, is very low . Thus, $\beta_{1a}$ should be lower compared to $\beta_{1b}$. We estimate $\beta_{1b} = 0.32833$ and $\beta_{1a} = 0.30$. However, dengue can become severe in the next infections thus we may have $\beta_{2b} = 0.164165$, $\beta_{3b} = 0.164165$, and $\beta_{4b} = 0.164165$. We also have $\beta_{2a} = 0.15$, $\beta_{3a} = 0.15$, and $\beta_{4a} = 0.15$. From the same literature, we also obtain the value for the death rate through infection given by $\gamma = 0.0000002$. Moreover, data from the Department of Health shows that the dengue deaths is more prevalent in the age group  0-9 years old. This would imply that, $\gamma_{1a}$ should be higher compared to $\gamma_{1b}$. We let $\gamma_{1b} = 2.0 \times 10^{-7}$ and  $\gamma_{1b} = 3.0 \times 10^{-7}$. In the next infections, dengue can become severe and may lead to more number of deaths . Thus, we let $\gamma_{2b}=\gamma_{3b} =\gamma_{4b}= 4.0 \times 10^{-7}$ and $\gamma_{2a} = \gamma_{3a}=\gamma_{4a}= 6.0 \times 10^{-7}$.

From \cite{dengue}, we have the birth rate $\mu$ equal to $8.5 \times 10^{-4}$ and death rate $\delta$ equal to $4.5 \times 10^{-4}$. Furthermore, we estimate the growth rate from age group $a$ to age group $b$ to be $\epsilon_1=6.92 \times 10^{-5}$ and the growth rate from age group $b$ to age group $a$ to be $\epsilon_2=6.92 \times 10^{-5}$.

\subsection{Numerical Simulations}
For our numerical simulations, we use the following initial values:\\
$S_{4a} =  2.1 \times 10^{7}$, \,
$I_{1a} =  7.0 \times 10^{6}$, \,
$S_{3a} =  8.0 \times 10^{6}$, \,
$I_{2a} = 4.0 \times 10^{6}$, \,
$S_{2a} =  4.5 \times 10^{6}$, \,
$I_{3a} = 2.5 \times 10^{6}$, \,
$S_{1a} = 1.0 \times 10^{6}$, \,
$I_{14a} =3.5 \times 10^{5}$, \,
$R_a=1.5 \times 10^{6}$, \,
$S_{4b} = 2.1 \times 10^{7}$, \,
$I_{1b} =  1.4 \times 10^{7}$, \,
$S_{3b} =  1.0 \times 10^{7}$, \,
$I_{2b} =  6.0 \times 10^{6}$, \,
$S_{2b} = 4.5 \times 10^{6}$, \,
$I_{3b} = 2.5 \times 10^{6}$, \,
$S_{1b} =  1.5 \times 10^{6}$, \,
$I_{4b} =  6.5 \times 10^{5}$, \, and
$R_b=3.5 \times 10^{5}$.

With the estimated parameters above and the set of initial conditions, and being assured of the existence of the optimal controls $u_1^{*}(t), u_2^{*}(t),$ and $u_3^{*}(t)$ by Theorem 4.1.1, we performed the numerical simulations. The results are given by the following figures. On each of the figures, the graph on the left shows how the control should be implemented to have the minimum value of the objective functional and the graph on the right shows the corresponding effect of the implementation on the total number of infected individuals. The black graph represents the total number of infected individuals over time when no control is used.    

\begin{figure}[H]
	\begin{center}
		\includegraphics[width=15cm,height=40cm, keepaspectratio]{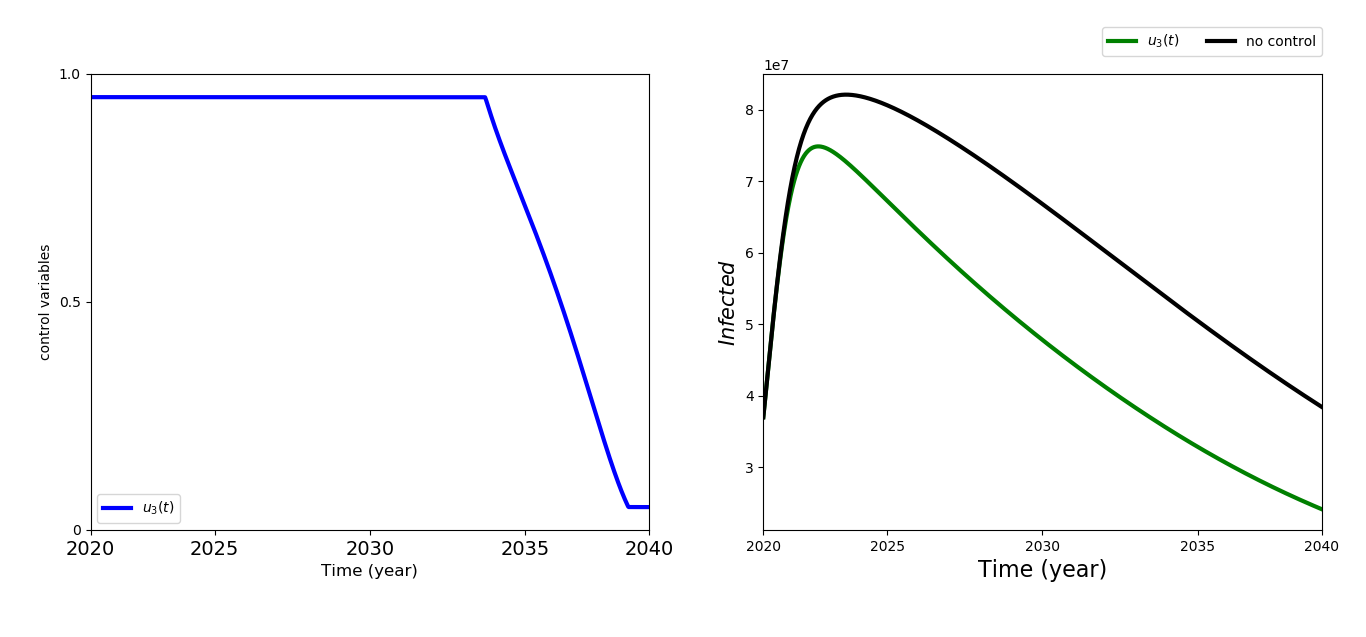}
	\end{center}
\vspace{-30pt}
	\caption{Optimal strategy and corresponding effect for Dengvaxia}
\end{figure}
 
\begin{figure}[H]
	\begin{center}
		\includegraphics[width=15cm,height=40cm, keepaspectratio]{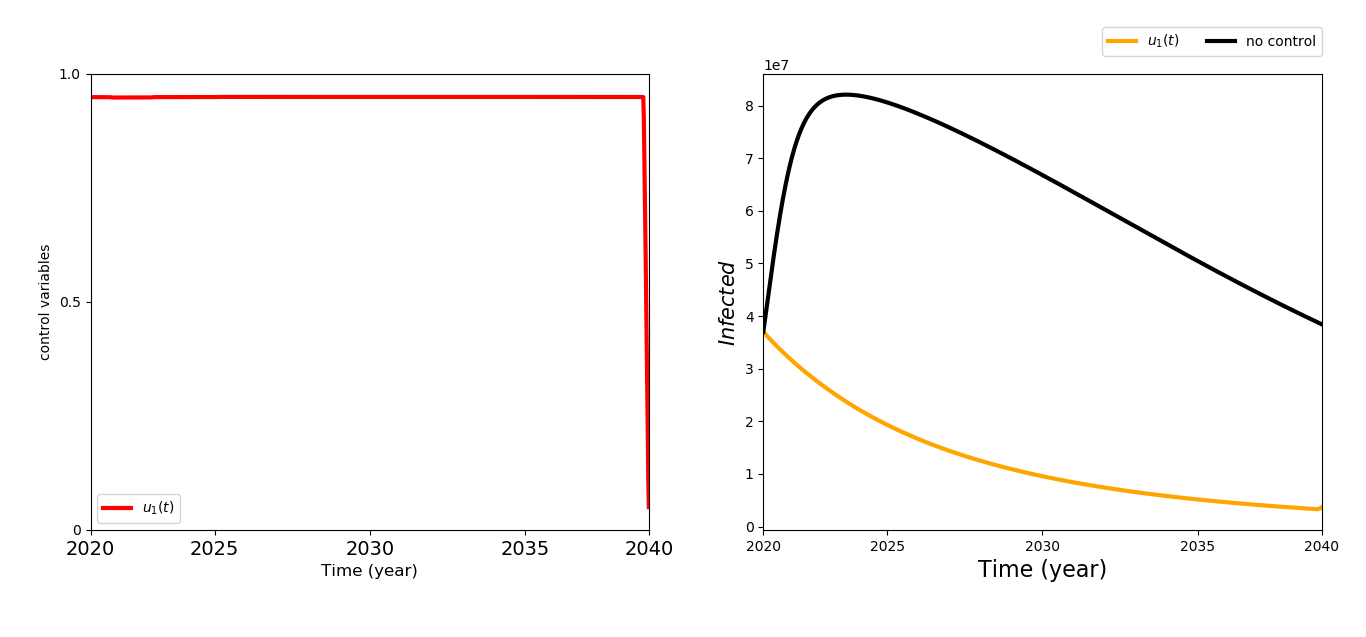}
	\end{center}
\vspace{-30pt}
	\caption{Optimal strategy and corresponding effect for tranmission reduction control}
\end{figure}
  
\begin{figure}[H]
	\begin{center}
		\includegraphics[width=15cm,height=40cm, keepaspectratio]{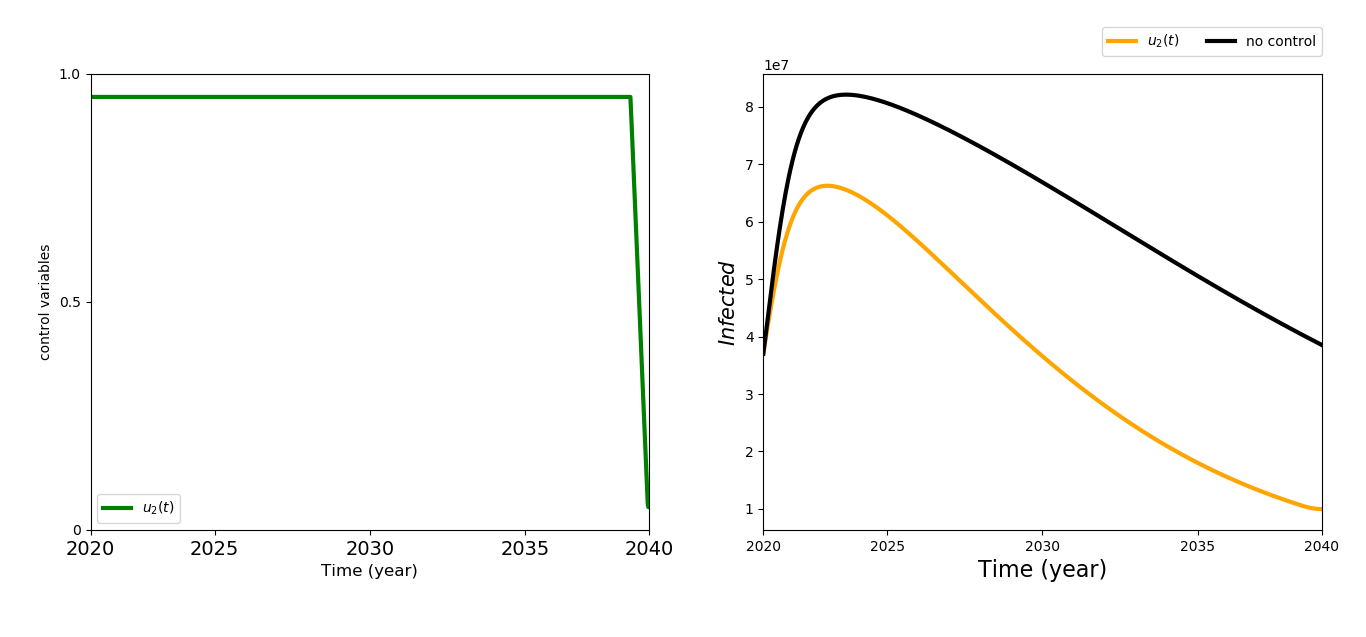}
	\end{center}
\vspace{-30pt}
	\caption{Optimal strategy and corresponding effect for proper medical care control}
\end{figure}

\begin{figure}[H]
	\begin{center}
		\includegraphics[width=15cm,height=40cm, keepaspectratio]{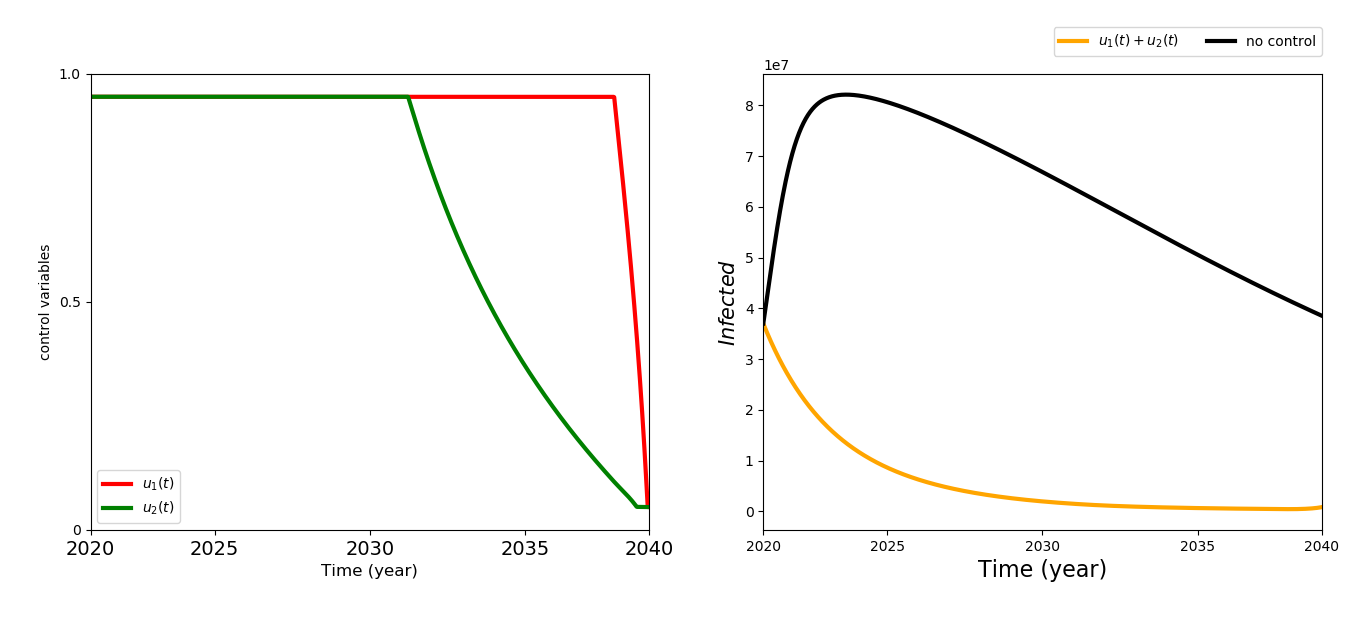}
	\end{center}
	\vspace{-30pt}
	\caption{Optimal strategy and corresponding effect for the combined transmission reduction and proper medical care controls}
\end{figure}

\subsection{Discussion}
As expected, we can see a significant decrease in the total number of infected individuals if Dengvaxia is implemented with the optimal strategy. However, one can clearly see that the optimal strategies for the other alternative controls have better impacts than that of Dengvaxia. The best impact if only one control is to be used can be seen in the optimal strategy for the transmission reduction control. These results is telling us that even without Dengvaxia we can still reduce the number of infected individuals and the reduction could be significantly better. It is interesting to note that the transmission reduction control is better than the proper medical care control, showing that prevention is still better than cure.

However, there is a very important point that should be made here. Note that the effects can only be seen if the optimal strategies are used. For the transmission reduction and proper medical care control, the graphs on the left shows us that the optimal strategy is almost maximum administration of the control for the duration of the program. This means that almost all of the affected population should do their part in implementing these strategies, if possible, at all times. This clearly calls for discipline, and this is where the Dengvaxia control has the advantage. The vaccination does not need discipline from the population to be implemented. Therefore, a concerted effort from the people and the government is needed to effectively control dengue without the aid of vaccination like Dengvaxia, and doing so seems rewarding because the expected effect seems to be way better than the effect of Dengvaxia and the implementation will not cause any Dengvaxia-like public fear.

\end{document}